\documentclass[twocolumn,showpacs]{revtex4}
\usepackage{graphicx}
\usepackage{bm,amsmath,amssymb}

\newcommand{\ds}{\Delta S}
\newcommand{\du}{\Delta U}
\newcommand{\dn}{\Delta N}
\newcommand{\de}{\delta E}
\newcommand{\dom}{\delta \omega}
\begin{document}
\title{Fluctuation theorem for black-body radiation}
\author{B. Cleuren}
\affiliation{Hasselt University - B-3590 Diepenbeek, Belgium}
\author{C. \surname{Van den Broeck}}
\affiliation{Hasselt University - B-3590 Diepenbeek, Belgium}
\begin{abstract}
The fluctuation theorem is verified for black-body radiation, provided the bunching of photons is taken into account appropriately.
\end{abstract}
\pacs{02.50.-r,05.40.-a,42.50.Ar,42.50.Lc}\maketitle

%==============================================================
% INTRODUCTION
%==============================================================
\section{Introduction}
The study of black-body radiation has played a prominent role in the discovery of quantum mechanics. More recently, the quantum features of light have been revealed in many beautifull optical experiments, and found to be in full agreement with the quantum theory of electromagnetic radiation  \cite{mandel,scully,paul}. In particular, we cite the bunching of photons. Bunching refers to the tendency of photons to  arrive together at closely spaced detectors, resulting in super-Poissonian counting statistics (variance large than the mean).  While a classical explanation is available \cite{purcell}, the detailed understanding and description of the phenomenon requires quantum field theory \cite{mandel,beenakker}.  The bunching of photons, and more generally of bosons  \cite{socialatoms}, is usually explained in terms of the more familiar quantum statistical exchange force which introduces an effective attraction between identical bosons, when their wavefunctions overlap \cite{griffiths}. The phenomenon of bunching was first observed by Hanbury Brown and Twiss \cite{HBT} as the correlation between intensities of light from a single star falling on different detectors. Bunching of photons in a single detector experiment was later observed \cite{mandelPRL} and found to be in agreement with a simplified quantum mechanical argument by Glauber \cite{qmarg}. For the case of equilibrium black-body radiation, the photon count has been calculated explicitly \cite{beenakker} and is given by a negative binomial distribution rather than the Poissonian distribution which describes independent arrivals.

In this Letter, we show that this result for the photon counting statistics is essential to find agreement with a recent result from nonequilibrium statistical mechanics, namely, the so-called fluctuation theorem. This theorem states that the probability distribution $P(\Delta S)$ to observe an entropy production $\Delta S$ during a time interval $t$ in a nonequilibrium steady state obeys the following symmetry relation for asymptotically long values of $t$ \cite{evans93,gallavotti95,kurchan,lebowitz,maes,evansREVIEW,derrida2}:
\begin{equation}\label{eq:ft}
\frac{P(\Delta S)}{P(-\Delta S)} \sim \exp\{\Delta S/k \}.
\end{equation}
In words, the probability of observing a positive entropy change is exponentially larger than that of the corresponding negative change. When the system starts in a state of canonical equilibrium which is subsequently perturbed by a time-dependent change of the Hamiltonian, the above fluctuation theorem, referred to as the transient fluctuation theorem, is valid for all times, and not just asymptotically large ones \cite{evansREVIEW,cleurenPRL}.\newline
The fluctuation theorem has been verified in several theoretical \cite{hatano,faragoJSP2002,jarzynskiPRL2004,derridaJSP2004,gaspard1,seifertPRL2005,seifertEPL2005,gaspard2,gilbertPRE2006,cleurenPRE2006,naudtsPRE2006,porporatoPRL2007,wood} and experimental settings \cite{wangPRL2002,ciliberto,garnierPRE2005,schuler,blickle2006,tietzPRL2006,andrieuxPRL2007}. The physical origin of the fluctuation theorem is to be found in the time-reversal symmetry of the underlying Hamiltonian dynamics. As such, this result can be viewed as a generalization of the ideas of Onsager \cite{astumian}.\newline 
We shall investigate the steady state of radiation exchange between two black bodies, at equilibrium at  different temperatures, when they are connected to each other through a small aperture. To present the problem and to contrast the role of the quantum mechanical bunching for photons with that of classical particle exchange, we start by studying the classical counterpart of effusion of the ideal gases.

%==============================================================
% IDEAL GAS
%==============================================================
\section{Effusion of a classical ideal gas}
\begin{figure}
\includegraphics[width=0.45\textwidth]{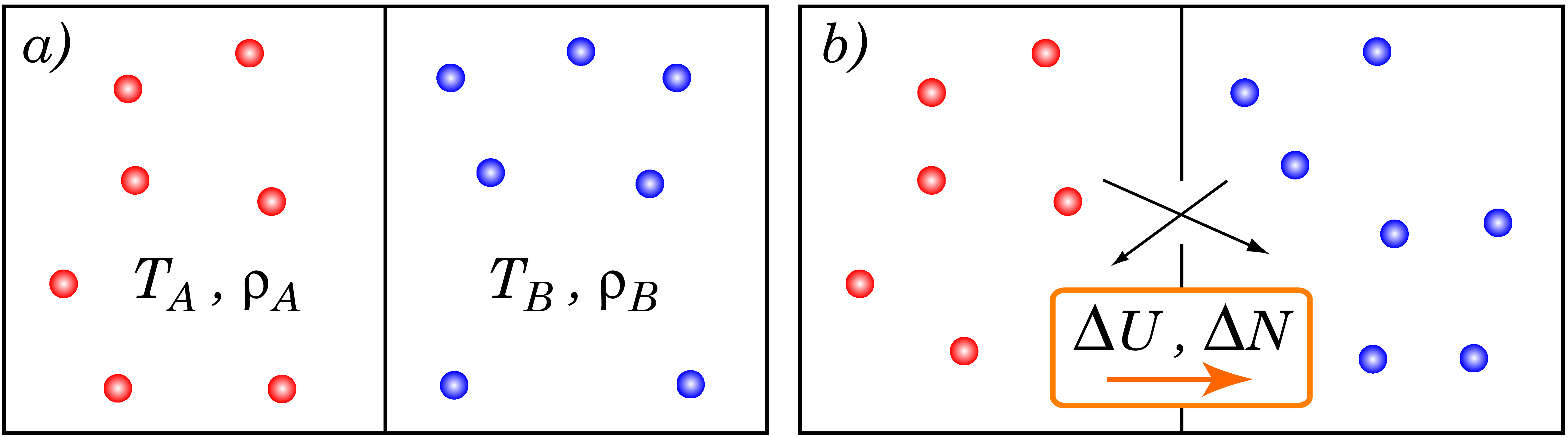}
\caption{(Color online) Set-up for the effusion of ideal gases: a) Two separate ideal gases in equilibrium at their respective temperature and density. b) During a time $t$, a small hole is opened and the net transfer of energy $\du$ and $\dn$ is measured.}
\label{figIGsetup}
\end{figure}

\begin{figure}
\includegraphics[width=0.20\textwidth]{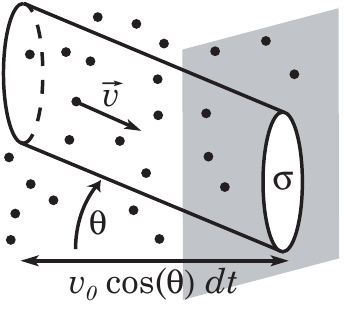}
\caption{The escape rate is determined by counting the particles that have the correct kinetic energy and which are able to reach the opening in a short time $dt$ (these are located in the cylinder with volume $\sigma v_{0}\cos(\theta)dt$, $v_{0}=\sqrt{2E_{0}/m}$). Inserting the Maxwellian velocity distribution and adding up all contributions from the different angles leads to the expression eq.~(\ref{eq:rate}).}
\label{figTR}
\end{figure}

Consider two infinitely large reservoirs $A$ and $B$, each containing classical ideal gases at  (canonical) equilibrium with temperatures $T_A$ and $T_B$ and densities $\rho_A$ and $\rho_B$, respectively. We perform the following experiment (see fig. \ref{figIGsetup}). During a fixed time interval $t$, a small hole of surface area $\sigma$ between the reservoirs is opened. We assume, for better comparison with the photon crossings, that the opening contains an energy filter, allowing only particles with kinetic energy in the range $E_{0}\pm \de/2$ to  move across the hole. The net amount of energy $\du$ and net number of particles $\dn$ transferred from $A$ to $B$ are measured.  We consider the limit of a small energy window, $\de \ll E_{0}$, so that $\du = E_{0}\dn$, and the exchange in the number of particles is the only relevant variable. Since the hole is small enough so that the canonical equilibria in the reservoirs are not significantly perturbed, the corresponding entropy change $\ds$  is given by standard thermodynamics:
\begin{equation}\label{eq:ds}
\ds = \left\{\left(\frac{1}{T_{B}}-\frac{1}{T_{A}}\right)E_{0}+\left(\frac{\mu_{A}}{T_{A}}-\frac{\mu_{B}}{T_{B}}\right)\right\}\dn.
\end{equation}
Inserting the well-known espressions \cite{callen} for the chemical potentials $\mu_{\alpha}$ of the ideal gases in reservoirs $\alpha \in \{A,B\}$, one finds:
\begin{equation}\label{eq:chemPot}
\frac{\mu_{A}}{T_{A}}-\frac{\mu_{B}}{T_{B}}=k\log\left(\frac{\rho_{A}}{\rho_{B}}\left[\frac{T_{B}}{T_{A}}\right]^{\frac{3}{2}}\right).
\end{equation}
\newline
Since the particle exchange $\dn$ is the result of individual gas particle crossings, it is obviously a random variable, and hence so is $\ds$. 
%------figure 1---------
%------figure 1---------
The  probability distribution $P(\dn)$  can be calculated as follows. We define $p_{\alpha}(j)$ as the probability that $j$ particles leave the reservoir $\alpha$ in the specified time interval. As all the particles move independently of one another, $p_{\alpha}(j)$ is Poissonian  \cite{vankampen}:
\begin{equation}\label{eq:pois}
p_{\alpha}(j)=\frac{(\nu_{\alpha}t)^{j}}{j!}e^{-\nu_{\alpha}t}.
\end{equation}
The escape rate $\nu_{\alpha}$ can be calculated following standard arguments from kinetic theory (see fig.~\ref{figTR}). We get
\begin{equation}\label{eq:rate}
\nu_{\alpha}=\frac{\sigma \rho_{\alpha}}{\sqrt{2\pi m k T_{\alpha}}}\frac{E_{0}}{k T_{\alpha}}e^{-E_{0}/kT_{\alpha}}\de.
\end{equation}
Since the escape of particles from the left and right reservoirs are independent phenomena, $P(\dn)$ is given by the sum
\begin{equation}\label{eq:som}
P(\dn)=\sum_{i=0}^{\infty}p_{A}(\dn+i)p_{B}(i).
\end{equation}
Combining eqs.~(\ref{eq:pois}) and (\ref{eq:som}) leads to the following explicit expression for $P(\dn)$:
\begin{equation}
P(\dn)=e^{-t(\nu_{A}+\nu_{B})}\left(\frac{\nu_{A}}{\nu_{B}}\right)^{\frac{\dn}{2}}I_{\dn}\left(2t\sqrt{\nu_{A}\nu_{B}}\right).
\end{equation}
Recalling that the modified Bessel function $I_{\dn}$ is an even function of $\dn$, the fluctuation theorem is verified by direct substitution of this result in eq.~(\ref{eq:ft}). 
\newline 
We mention a simpler proof, more clearly related to the underlying micro-reversibility. Consider the exchange process, with $\dn+i$ particles going from $A \rightarrow B$, and $i$ particles from $B \rightarrow A$. This occurs with probability $p_{A}(\dn+i)p_{B}(i)$. The time reversed process, with $i$ particles going from $A \rightarrow B$ and $\dn+i$ particles from $B \rightarrow A$, has probability $p_{A}(i)p_{B}(\dn+i)$. Their ratio  satisfies the following detailed fluctuation theorem:
\begin{multline}\label{eq:dft}
\frac{p_{A}(\dn+i)p_{B}(i)}{p_{A}(i)p_{B}(\dn+i)}=\left(\frac{\nu_{A}}{\nu_{B}}\right)^{\dn}\\=\left(\frac{\rho_{A}}{\rho_{B}}\left[\frac{T_{A}}{T_{B}}\right]^{\frac{3}{2}}\exp\left\{\left[\frac{1}{T_{B}}-\frac{1}{T_{A}}\right]\frac{E_{0}}{k}\right\}\right)^{\dn}
=e^{\frac{\ds}{k}},
\end{multline}
using eqs.~(\ref{eq:rate}) and (\ref{eq:ds}). The fluctuation theorem itself follows immediately. We have
\begin{eqnarray}\label{pft}
P(\dn)&=&\sum_{i=0}^{\infty}p_{A}(\dn+i)p_{B}(i) \nonumber \\&=&\sum_{i=0}^{\infty}\left(\frac{\nu_{A}}{\nu_{B}}\right)^{\dn}p_{A}(i)p_{B}(\dn+i)
\nonumber \\ &=&\left(\frac{\nu_{A}}{\nu_{B}}\right)^{\dn}P(-\dn).
\end{eqnarray}
%------figure 2---------
%------figure 2---------

%==============================================================
% PHOTONS
%==============================================================
\section{Black-body radiation}
\begin{figure}
\includegraphics[width=0.45\textwidth]{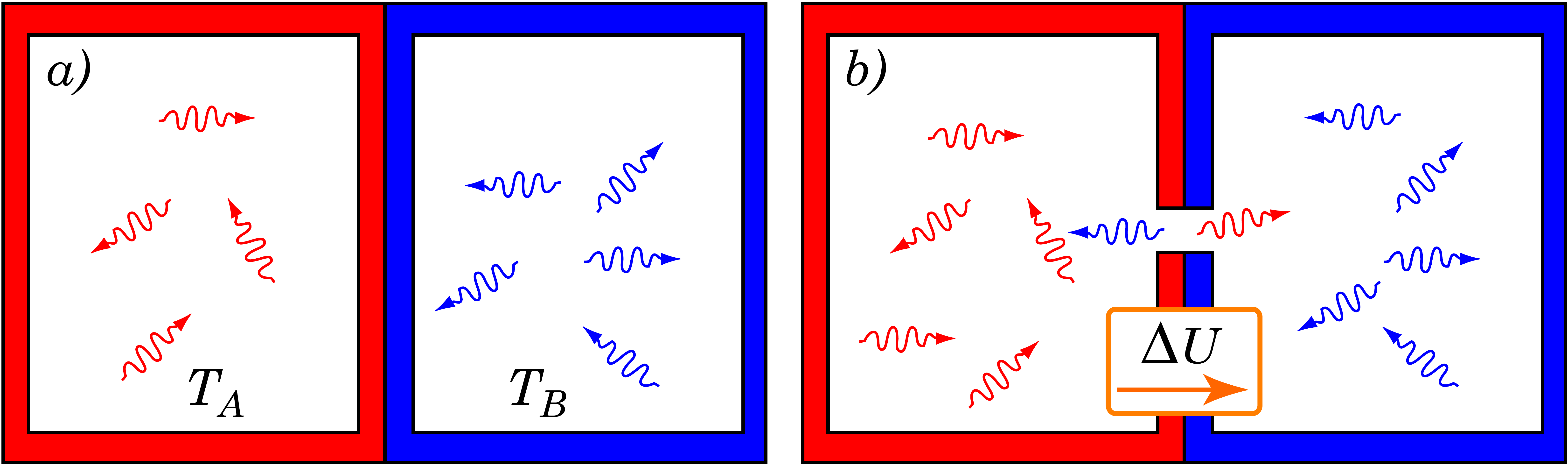}
\caption{(Color online) Set-up for black-body radiation: a) Two cavities are equilibrium each at its own temperature. b) During a time $t$, a small hole is opened and the net transfer of energy $\du$ is measured.}
\label{figPHsetup}
\end{figure}

We now turn to a similar set-up for the study of fluctuations in black-body radiation. Two large empty cavities ($A$ and $B$), whose walls are kept at a fixed, but different temperatures, $T_{A}$ and $T_{B}$ respectively, act as  sources of black-body radiation (see fig. \ref{figPHsetup}). During a fixed time interval $t$, radiative exchange becomes possible by opening a small aperture (surface area $\sigma$), which permits free passage of photons with a frequency in the range $\omega_{0}\pm \dom/2$. We again consider the limit of a monochromatic filter  $\dom \ll \omega_{0}$.  Following standard thermodynamics, the entropy change $\ds$, upon transfer of a net number of photons $\dn$ from $A$ to $B$ (and hence of energy  $\hbar \omega_{0} \dn$), is given by:
\begin{equation}\label{eq:entropy}
\ds = \left(\frac{1}{T_{B}}-\frac{1}{T_{A}}\right)\hbar \omega_{0} \dn.
\end{equation}
Photons coming from the two different reservoirs are independent of each other. Hence the probability $P(\dn)$ to observe a net transfer of $\dn$ photons from $A$ to $B$ is again given by eq.~(\ref{eq:som}). However, photons coming from the same reservoir are not independent, and their escape is no longer governed by a Poisson distribution. The calculation of $p_{\alpha}(j)$, which is also referred to as the photon counting distribution, requires a fully quantum mechanical description of the electromagnetic field \cite{mandel,beenakker}. For large times $(t\dom \gg 1)$, it is given by the following negative binomial distribution:
\begin{equation}
p_{\alpha}(j)=\frac{\Gamma (j+\nu t)}{j!\Gamma(\nu t)}\left(1-e^{-\hbar \omega_{0/kT_{\alpha}}}\right)^{\nu t}\left(e^{-\hbar \omega_{0/kT_{\alpha}}}\right)^{j},
\end{equation}
where $\nu$ is defined as
\begin{equation}
\nu = \frac{\sigma \omega_{0}^{2} \dom}{(2\pi c)^{2}}.
\end{equation}
It is now a matter of simple algebra to verify that the following detailed fluctuation theorem is obeyed:
\begin{equation}\label{eq:ratiodft}
\frac{p_{A}(\dn+i)p_{B}(i)}{p_{A}(i)p_{B}(\dn+i)}= e^{\left(\frac{1}{T_{B}}-\frac{1}{T_{A}}\right)\hbar \omega_{0} \dn/k}.
\end{equation}
Consequently, following the steps of eq.~(\ref{pft}), the fluctuation theorem is also verified.\newline
We emphasize that the correlations between the photons, resulting in the negative binomial distribution, is an essential ingredient. A semiclassical approach, which assumes statistical independence between photons, cannot be reconciled with the fluctuation theorem (see Appendix). 

%==============================================================
% DISCUSSION
%==============================================================
\section{Discussion}
We close with a number of comments on the above results.\newline
We have presented the analysis above for the case of a monochromatic window connecting the reservoirs. This permits the derivation of a detailed fluctuation theorem. It is also revealing to note that the entropy production becomes identically zero for the classical gas for a particular choice of the energy window: the r.h.s. of eq.~(\ref{eq:dft}) is identically equal to unity for a particular value of the energy, namely,
\begin{equation}
E_0=k \frac{T_A T_B}{T_A-T_B} \ln \left\{\frac{\rho_{A}}{\rho_{B}}\left[\frac{T_{A}}{T_{B}}\right]^3\right\}.
\end{equation}
At this specific value of the kinetic energy, the one-particle energy distributions in reservoirs $A$ and $B$ cross each other, so that the particles with this energy have the same number density on both sides.  Such an equilibrium state can be used to reach optimal thermodynamic Carnot and Curzon-Ahlborn efficiencies \cite{vandenbroeck}. This type of equilibrium can also be achieved with electrons \cite{linke}, but not with photons because the latter have zero chemical potential.\newline
The fluctuation theorem for the effusion of an ideal gas is valid for all times, while for photons the theorem has been proven here only for large times. This is consistent with the observation that the effusion of an ideal gas is a process without memory. In such a case, the distinction between the steady state and transient versions of the fluctuation theorem disappears. A transient theorem for photons valid for all times can also be obtained in principle, but this would require the exact evaluation of the transient photon count, starting from zero and progressing to the steady state, after the aperture is opened.\newline
As expected from the general argument connecting  bunching to the statistical exchange interaction, one finds that fermions display anti-bunching \cite{jeltes}. While this phenomenon has no classical analogue, it is surprising to find, for example, that the statistics of an electron current through a small channel can be fully reproduced by a simple classical random walk model, namely, the symmetric exclusion process \cite{derridaJSP2004}. For this simple model, the fluctuation theorem is again verified. The validity of  the fluctuation theorem for fermions has also been confirmed by quantum field theoretic calculations \cite{esposito2}.

%==============================================================
% APPENDIX
%==============================================================
\section{Appendix}
A semiclassical description assumes the photons to be independent of each other. The photon counting distribution is then given by eqs.~(\ref{eq:pois}) and (\ref{eq:som}). In this case, the quantum aspects of the photons are only taken into account via their occupation number density as implied by Planck's law of radiation. This leads to the following result for the escape rate:  
\begin{equation}
\nu_{\alpha}=\frac{\sigma \omega_{0}^{2}\dom}{(2\pi c)^{2}\left(e^{\hbar \omega_{0}/kT_{\alpha}}-1\right)}.
\end{equation}
The ratio of these rates (cf. eq.~(\ref{eq:dft})) does not yield the desired result $\exp\{\ds/k\}$ with $\ds$ given by eq.~(\ref{eq:entropy}), and hence the fluctuation theorem is not satisfied.\newline
The picture of independent photons has been used to calculate the  average number of photons leaving compartment $\alpha$ in a time interval $t$ (see, e.g., \cite{dorfman}). Since this quantity is not influenced by bunching, the correct result is obtained in this case:
\begin{equation}
\langle j \rangle = \nu_{\alpha} t.
\end{equation}
Integrating over $\dom$, one recovers the Stefan-Boltzmann law. However, the higher order moments are not correctly reproduced in the independent photon picture.
%==============================================================
% BIBLIOGRAPHY
%==============================================================

\end{document}